﻿

\documentclass{jkas}

\def\year{2020} 
\def\month{February} 
\def\volume{00} 
\def\issue{0} 
\def\beginpage{1} 
\def\endpage{99} 
\def\received{February ??, 2020} 
\def\accepted{February ??, 2020} 
\date{Received \received; accepted \accepted}

\title{
Poorly Studied Eclipsing Binaries in the Field of DO Dra: V0455 Dra and V0454 Dra 
}

\author[1]{Yonggi Kim
}
\author[2]{Ivan L. Andronov}
\author[2]{Kateryna D. Andrych}
\author[1]{Joh-Na Yoon}
\author[1]{Kiyoung Han}
\author[3,2]{Lidia L. Chinarova}

\affil[1]{Chungbuk National University Observatory, Chungbuk National University, 361-763, Cheongju, Korea \email{ykkim153@chungbuk.ac.kr, antalece@chungbuk.ac.kr}}
\affil[2]{Department "Mathematics, Physics and Astronomy", Odessa National Maritime University, 65029, Odessa,  Ukraine; \email{tt{\_}ari@ukr.net, katyaandrich@gmail.com}}
\affil[3]{Astronomical Observatory, Odessa I.I.Mechnikov National University, 65014, Odessa, Ukraine; \email{llchinarova@gmail.com}
}

\def\corrauthor{
Yonggi Kim
}

\def\runningauthor{
Kim Yonggi et al.
}

\def\runningtitle{
Eclipsing Binaries V0454 Dra and V0455 Dra
}

\def\keywords{
(stars:) binaries: eclipsing;
methods: data analysis;
stars: individual: V0455 Dra -- V0454 Dra -- DO Dra -- GSC 04395-00485 -- GSC 04396-00605 --  V1517 Her
}

\def\abstracttext{
We report an analysis of two poorly studied systems GSC 04396-00605 and GSC 04395-00485, which were recently named as V0455 Dra and V0454 Dra, respectively. For two eclipsing stars, the periods and epochs were significantly corrected using our extensive data. The phenomenological characteristics of the mean light curves were determined using the “New Algol Variable” (NAV) algorithm.  The individual times of Maxima/Minima (ToM) were determined using the newly developed software MAVKA, which outputs the accurate parameters using the ''asymptotic parabola'' approximations. The light curves were approximated using the Wilson-Devinney model, the best fit parameters are listed.
For both systems, the presence of the dark spot is justified, the parameters are estimated.
Both systems classified to be overcontact binaries of EW type.
The observations were obtained using the 1-m Korean telescope of the LOAO while monitoring the cataclysmic variable DO Dra within the “Inter-Longitude Astronomy" (ILA) project.
}

\begin{document}
\jkashead 


\section{Introduction\label{sec:intro}}

All the stars become variable at some active stages of their evolution. Of the special interest, there are objects exhibiting few types of variability, e.g. cataclysmic variables and other interacting binaries, pulsating and eruptive variables.
The photometrical monitoring of such objects needs many dozens or even hundred of nights, and may not be done at very large telescopes. 
This is a proper task for meter and sub-meter class telescopes. There are single- and multi-telescope projects on some selected stars. Among these projects, there is an international campaign ILA (''Inter-Longitude Astronomy'', \cite{a03}). 
The most recent review on highlights was presented by \cite{a17a}. 

One of the most interesting stars in our sample is DO Dra, which may represent a separate group of objects, being simultaneously the ''magnetic dwarf nova'' and the ''outbursting intermediate polar''. It has shown a variety of new phenomena like the ''transient periodic oscillations'' (TPO), dependence of the slope of the outburst decay on the brightness at maximum, a weak wave of the ''basic low'' brightness between the outbursts, many of which seem to be missing because of a short duration (\cite{a08}). A statistical study of the QPOs was presented by \cite{k17}. 

Besides this very interesting object, there are other variables in the field. Two of them were discovered in a frame of the ILA project by \cite{v10}  while searching for variability of stars on various fields. \cite{v11} reported on new variables in the field of DO Dra, presented finding charts and the preliminary parameters. The systems were classified as the $\beta$ Lyr-type (EB) and W UMa-type (EW) objects. Recently, the stars got the official names V0454 Dra and V0455 Dra in the "General Catalogue of Variable Stars" (GCVS) (\cite{k15}, \cite{s17}). Names of the systems in other catalogues are 
 2MASS J11403001+7111021= GSC  4395.00485 = USNO-B1.0 1611-00091333 = USNO-B1.0 1611-00091333 Gaia DR2 1062691908235241216 = V0454 Dra = V454 Dra, and
2MASS J11483649+7107507 = NOMAD-1 1611-0093251 = USNO-B1.0 1611-0091801 = V0455 Dra = V0455 Dra.

The parallaxes for these stars are $0.7722\pm0.0201$ mas and $0.8807\pm0.0186$ mas (\cite{gaia18a},\cite{gaia18b}), respectively.


\section{Observations}
Photometric observation data of DO Dra had been obtained
from Mt. Lemmon optical astronomy observatory (LOAO) and
Chungbuk national university observatory (CBNUO) for about
10 years from 2005 till 2014. The LOAO telescope located at
Mt. Lemmon in Arizona is 1.0 m in diameter and the effective
focal ratio is f/7.5. It is mounted on a fork equatorial mount and
adopts CCD with 0.64 arcsec/pixel, $2k \times 2k$ resolution, and its
field of view (FOV) is ${22'}.2  \times {22'}.2$. The telescope at CBNUO is 0.6
m in diameter and its optical system is an R-C type with effective
focal ratio of f/2.92. in 2012, a CCD with wide FOV of ${72'} \times {72'}$
and $4k \times 4k (4096 \times 4096)$
 pixel resolution was been installed.
The observations were made using the R filter, using the comparison star "Ref 2" from \cite{v11} and adopting its magnitude $Rc=13.04^m$ $(V=13.390^m,$ $V-R=0.350^m)$
(\cite{h07}).

Totally, we have analyzed $n=1746$ observations obtained during totally 112 hours distributed in 35 nights from JD 2456272 to 2456710 (total duration of the interval 438$^{\rm d}$).

\section{Phenomenological Modelling\label{sec:phen}}

Phenomenological modelling of the light curves of eclipsing systems using ''special shapes'' (also called ''patterns''), instead of trigonometrical polynomials, was proposed by \cite{a12} initially for the Algol-type (EA) systems with very clear begin and end of the eclipse. Thus the algorithm was called NAV (''New Algol Variable''). 

However, it is effective also for EB $(\beta$ Lyr) and 
EW (W UMa) type systems, also for the prototypes of these classes - $\beta$ Per (Algol), $(\beta$ Lyr) and EW UMa (\cite{t16}).

For another eclipsing binary 2MASS J18024395 + 4003309 = VSX J180243.9+400331 (currently named as V1517 Her) in the field of 
the intermediate polar V1323 Her, \cite{a15} estimated physical parameters of the components, using two-color photometry and a statistical mass-radius-color index relation.

For the stars V0454 Dra and V0455 Dra, such a complete analysis is not possible, as the photometry is in one color only.

We have computed periodograms using trigonometrical polynomial (TP) approximations of various orders $s.$ Then the periods were corrected using differential equations (\cite{a94}) according to the algorithm described by \cite{a94} and \cite{am06}. The statistically optimal values degrees of the trigonometrical polynomial are $s=6$ and $s=8$ for V0454 Dra and V0455 Dra, respectively. The corresponding number of parameters are $m=2s+2,$ i.e. $m=14$ and 18, more than in the NAV algorithm even in this case of relatively smooth curves.

The phenomenological parameters obtained using the NAV algorithm are listed in Table \ref{tab:jkastable1}. Their detailed description may be found in \cite{a15} and \cite{t16}.

\begin{table*}[t]
\caption{Parameters of the phenomenological modelling \label{tab:jkastable1}}
\centering
\begin{tabular}{lrr}
\toprule
Parameter & V0454 Dra & V0455 Dra \\ 
\midrule
$C_1$ & 13.5955 $\pm 0.0011^m$         &            14.2227$\pm 0.0022^m$\\
$C_2$ &  0.0061 $\pm  0.0011^m$         &            0.0299 $\pm 0.0021^m$ \\
$C_3$ & -0.0047 $\pm  0.0006^m$         &            0.0032 $\pm 0.0012^m$\\
$C_4$ &  0.1064 $\pm 0.0017^m$         &            0.1259 $\pm 0.0032^m$\\
$C_5$ &  0.0022 $\pm 0.0015^m$         &            0.0054 $\pm 0.0021^m$          \\
$C_6$ &  0.1276 $\pm  0.0042^m$         &            0.3101 $\pm 0.0088^m$ \\
$C_7$ &  0.1326 $\pm  0.0040^m$         &            0.2464 $\pm 0.0075^m$\\
$C_8$ &  0.1164 $\pm$  0.0025         &            0.1159 $\pm$ 0.0025          \\
$C_9$ &  1.369  $\pm$  0.079          &            1.357 $\pm$ 0.068\\
$C_{10}$&  1.539  $\pm$  0.081          &            1.330 $\pm$ 0.069\\
\addlinespace
$T_0$ &2456480.04281 $\pm  0.00030^d$ & 2456479.05227 $\pm 0.00023^d$       \\
$P$   &0.43491412  $\pm 0.00000091^d$  &       0.37683317   $\pm   0.00000097^d$\\
$d_1$ &    0.1109 $\pm$ 0.0034        &            0.2484 $\pm$   0.0061        \\
$d_2$ &    0.1149 $\pm$ 0.0033        &             0.2031$\pm$   0.0055        \\
$d1+d2$&   0.2258 $\pm$ 0.0058        &             0.4515$\pm$   0.0099        \\
$d1/d2$&   0.9646 $\pm$ 0.0291        &             1.2235$\pm$   0.0326        \\
Max I&   13.4845  $\pm 0.0012^m$        &            14.0999$\pm   0.0020^m$\\
Min I&   13.8355  $\pm 0.0032^m$        &            14.6886$\pm   0.0065^m$\\
Min II& 13.8284   $\pm 0.0027^m$        &            14.5651$\pm   0.0055^m$ \\
\bottomrule
\end{tabular}
\end{table*}

\section{Times of Extrema\label{sec:tom}}

For further studies of the period changes, it is important to publish individual times of extrema (e.g. \cite{k01}). Typically, only the moments of minima are published for eclipsing binaries. However, for both stars, the maxima are prominent, so we determined the maxima as well.

In the previous section, we modelled the complete phase curve based on all observations. In individual nights, the complete length of the observational run is smaller than the period. Thus algorithms should take into account only the intervals close to an extremum.

We have determined the times of either minima, or maxima using the ``running parabola'' approximations, initially proposed by \cite{ma96} and realized in the software MAVKA (\cite{aac15}, \cite{aac20}). This is one among the best (in accuracy) methods for determination of extrema from a relatively short intervals of observations near extremum, which only partially covers the descending and ascending branches. The ''special shapes'' allow to avoid apparent waves, which are similar to the Gibbs phenomenon. 

For longer series of observations, which completely cover ascending and descending branches, one may propose many modifications of known approximations. \cite{a17b} compared almost 50 functions and ranged them according to the accuracy estimate. Some of them are improvements of phenomenological approximations proposed by \cite{m15}. \cite{aa19} realized 21 approximations of 11 types for the shorter intervals.

In \ref{tab:jkastable2}, the moments of the individual minima and maxima are listed. To spare place, we have not marked the type of extremum. It is obvious that the moments, which correspond to phases near 0 and 0.5, are main (min I) and secondary (min II) minima, and the rest are the maxima.  

These moments may be used in further analysis, after adding further observations from other seasons, as typical timescale of the period variations are from year(s) due to third bodies, and much slower for the mass transfer (e.g. \cite{k01}).
\begin{table*}[t]
\caption{Moments of individual maxima and minima of V0454 Dra and V0455 Dra\label{tab:jkastable2}}
\centering
\begin{tabular}{lrlrlr}
\toprule
BJD-2400000 & $\pm$ &BJD-2400000 & $\pm$&BJD-2400000 & $\pm$\\ 
\midrule
{\bf V0454 Dra}&        & 56434.04771 & 0.0049 & 56394.07646 & 0.0007\\
56272.36895 & 0.0006 & 56650.31266 & 0.0015 & 56394.16601 & 0.0005\\
56346.30781 & 0.0006 & 56654.33204 & 0.0008 & 56395.11733 & 0.0007\\
56355.22405 & 0.0008 & 56658.35803 & 0.0013 & 56401.04896 & 0.0010\\
56357.07647 & 0.0016 & 56659.33758 & 0.0007 & 56409.06225 & 0.0010\\
56357.18191 & 0.0008 & 56667.27487 & 0.0006 & 56411.97503 & 0.0019\\
56357.28783 & 0.0013 & 56686.30513 & 0.0017 & 56412.07457 & 0.0006\\
56360.22388 & 0.0016 & 56710.22508 & 0.0019 & 56413.00769 & 0.0013\\
56362.18030 & 0.0003 & 56710.32959 & 0.0011 & 56413.10723 & 0.0011\\
56362.29182 & 0.0012 & {\bf V0455 Dra}&        & 56429.03185 & 0.0026\\
56363.26942 & 0.0009 & 56272.35864 & 0.0009 & 56434.02193 & 0.0009\\
56366.20593 & 0.0007 & 56346.31053 & 0.0019 & 56650.32477 & 0.0043\\
56373.05427 & 0.0013 & 56355.26561 & 0.0015 & 56651.36037 & 0.0014\\
56394.14774 & 0.0006 & 56357.06032 & 0.0012 & 56654.37722 & 0.0030\\
56395.12243 & 0.0015 & 56357.14830 & 0.0005 & 56658.23820 & 0.0015\\
56398.16881 & 0.0015 & 56357.23665 & 0.0010 & 56658.32863 & 0.0010\\
56401.10329 & 0.0012 & 56357.33556 & 0.0040 & 56659.27614 & 0.0012\\
56409.04061 & 0.0020 & 56360.16278 & 0.0009 & 56659.36724 & 0.0011\\
56411.97798 & 0.0020 & 56362.23293 & 0.0005 & 56664.35500 & 0.0035\\
56412.08446 & 0.0013 & 56363.17623 & 0.0008 & 56667.28187 & 0.0007\\
56413.06679 & 0.0006 & 56363.26870 & 0.0020 & 56686.31136 & 0.0005\\
56429.04618 & 0.0020 & 56366.18734 & 0.0019 & 56710.23717 & 0.0024\\
56432.98510 & 0.0026 & 56373.06110 & 0.0017 & 56710.33230 & 0.0032\\
\bottomrule
\end{tabular}
\end{table*}

\section{Physical Modelling\label{sec:phys}}

The main purpose for the all studies of eclipsing binaries data is to build physical models.  In contrast to the short set of parameters that can be determined for spectroscopic, visual or astrometric binaries, for eclipsing variables it is possible to determine masses, luminosities, sizes, temperatures, surface brightness distributions and some parameters of component orbits. 
For modeling and determination for these parameters we should combine photometric observations, radial velocity curves and spectroscopic observations.

The algorithm was proposed by \cite{wd71} and improved by \cite{w79} and \cite{w94}. The WD algorithm allows to determine physical parameters of binary system due to the phase curve and radial velocity curves. For this task, the user has to pre-set the intervals of the physical parameter values and to fix some of them for reasons related to the typical relationships for the stars of corresponding type of variability.

E.g. the temperature of the first component at the pole $T_1$  was fixed to a reasonable value, as typically assumed in the models, whereas the  temperature of the second component $T_2$ is a parameter being determined. The best fit estimate of $T_2$ is dependent on $T_1$ monotonically, but not exactly linearly. The surface potential $\Omega$ is expressed in units of square of the orbital velocity. As both stars overfill their Roche lobes and are at an overcontact stage, the potential is the same for the common envelope, i.e. for both stars. More details may be found in the monograph by \cite{km09}.

The algorithm is based on a geometry Roche (Roche model) for the construction of a binary system, and uses the Monte-Carlo method
to specify physical parameters and construct physical light curve. In this case, the WD-code allows to simulate different types of eclipsing systems from contact system to exoplanet transit.

We used the program to determine the physical parameters of the system, that is a modified version of the original WD-code created by Prof. S. Zola et al. at the Astronomical Observatory of Jagiellonian University in Krakow (\cite{z97}, \cite{z10}. This version can take into account up to two spots on components. 
The differential code method in the original code has been replaced with the Price algorithm (controlled Monte Carlo method). The version we use involves concretely the contact system, when both objects fill the Roche lobes and they both have the same potential on the surface, although they may have different temperatures (the temperature of one of them must be fixed while the other is simulated). Such a system is already tight, the components are tidal deformed and have circular orbits.

The co-latitude of the center of the spot varies from $0^\circ$ at the north pole of the star to $180^\circ$ at the south one. The spot center's longitude varies from $0^\circ$ to $360^\circ ,$ starting from the center line of the stars counter-clockwise.

Using the version of the original WD code created at the Jagiellonian Astronomical Observatory in Krakow by Prof. S. Zo{\l}a and others, we did a physical simulation of the eclipsing binary systems V0455 Dra and V0454 Dra. For these objects, we have only the phase curve in one filter, so some parameters and their possible intervals we fix as they “can be”. Physical parameters of components in close binary systems were discussed e.g. by \cite{k03}

The light curves for both objects obtained from the simulation are shown in Fig. 3 and 4 along with observations.
The resulting physical parameters are shown in Table \ref{tab:jkastable2}.

It should be noted that this is not a unique solution. There is a region in the parameter space, where the quality of the approximation is almost the same. There are pairs of parameters, which can compensate the influence of another one significantly, e.g. the temperatures of both components, the inclination and the potential, the radius of the spot and the temperature parameter.

\begin{table*}[t]
\caption{Parameters of the physical modelling \label{tab:jkastable3}}
\centering
\begin{tabular}{lrr}
\toprule
Parameter & V0454 Dra & V0455 Dra \\ 
\midrule
Orbit inclination & $74.67\pm 0.06^{\circ}$ & $64.\pm 0.03^{\circ}$\\ 
Temperature of the first star & 5800 K & 5500 K \\ 
Temperature of the second star & $5351\pm 5$ K & $5478\pm 3$ K \\ 
Surface potential & $3.62\pm 0.015$  & $3.685 \pm 0.0007$\\ Mass ratio & $0.95\pm0.01$  & $0.99\pm 0.005$ \\ 
Co-latitude of the spot's center & $90^{\circ}$ & $133\pm1.3^{\circ}$ \\ 
Longitude of the spot`s center & $356.9\pm 0.2^{\circ}$ & $71\pm 1.3^{\circ}$ \\ 
Spot radius & $22.5\pm 0.2^{\circ}$ & $18.9 \pm 0.74^{\circ}$ \\ 
Spot temperature (in the star temperature units) & 0.8 & 0.6 \\ 
\bottomrule
\end{tabular}
\end{table*}

With Binary Maker 3 \cite{bra05}, we built a visual model of the system with the appropriate physical parameters. In Fig. 5 and Fig. 6, the respective system models are shown. The phases are 0.24 and 0.74, where the spot and both components are seen. The red crosses represent the centers of stars and the positions of the center of mass of the system. The red circles indicate the orbits of the components around the common center of mass. To build the model, 120 points per meridian and 240 points per parallel are taken for each star. 

\section{Conclusions\label{sec:con}}

Photometric series for two poorly known stars V454 Dra and V455 Dra were analysed using phenomenological approximations of the complete phase light curve (using the trigonometrical polynomial and the NAV algorithms), as well as individual series near maxima and minima (using the ''asymptotic parabola'' algorithm implemented in a recently developed software MAVKA), the corresponding database of almost 70 moments was compiled. It may be used after years of further monitoring, to look for possible period variations.

The NAV algorithm shows better approximation in a physical sense, as the minima are modelled using the special shape, instead of the trigonometrical polynomial. The eclipses split the phase curve into four parts, so the smooth approximation using the trigonometrical polynomial is not justified physically.

These observations were also used for physical modelling according to the Wilson-Devinney algorithm.
The set of physical parameters, which were determined, may be used as an initial approximation, which may be improved using possible further spectral observations (to determine the mass ratio and temperatures) or well-calibrated multi-color photometry (to determine temperatures).
For both systems, the dark spot is evident, the parameters are estimated.


\acknowledgments

The authors thank Prof S. Zo{\l}a and Dr. B. D\c{e}bsky for allowing to use the software for physical modelling and fruitful discussions.
This work was supported by a research grant from
Chungbuk National University in 2017. The data acquisition
and analysis was partially supported by the Basic
Science Research Program through the National Research
Foundation of Korea (NRF) funded by the Ministry of
Education, Science and Technology (2011-0014954). It is
also a part of the “Inter-Longitude Astronomy” (Andronov et al. 2003, 2017a) and "Astroinformatics" (Vavilova et al. 2017) campaigns.

\begin{figure}
\centering
\includegraphics[width=80mm]{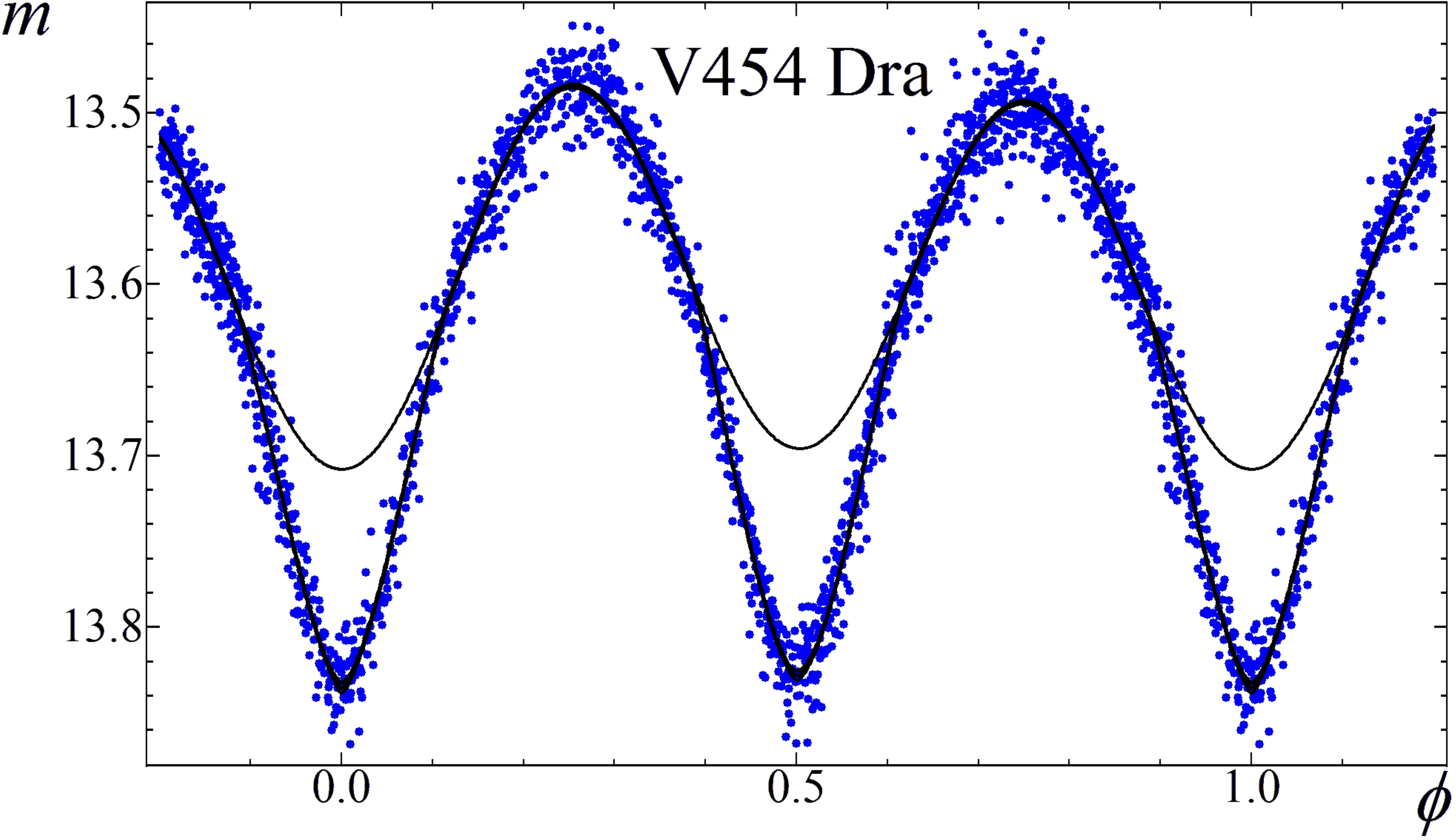}
\caption{The phase light curve of V0454 Dra and it's approximation using the NAV algorithm. A smooth line above the eclipse corresponds to the ''out-of-eclipse'' continuum apptoximated by a trigonometrical polynomial of order 2.\label{fig:jkasfig1}}
\end{figure}

\begin{figure}
\centering
\includegraphics[width=80mm]{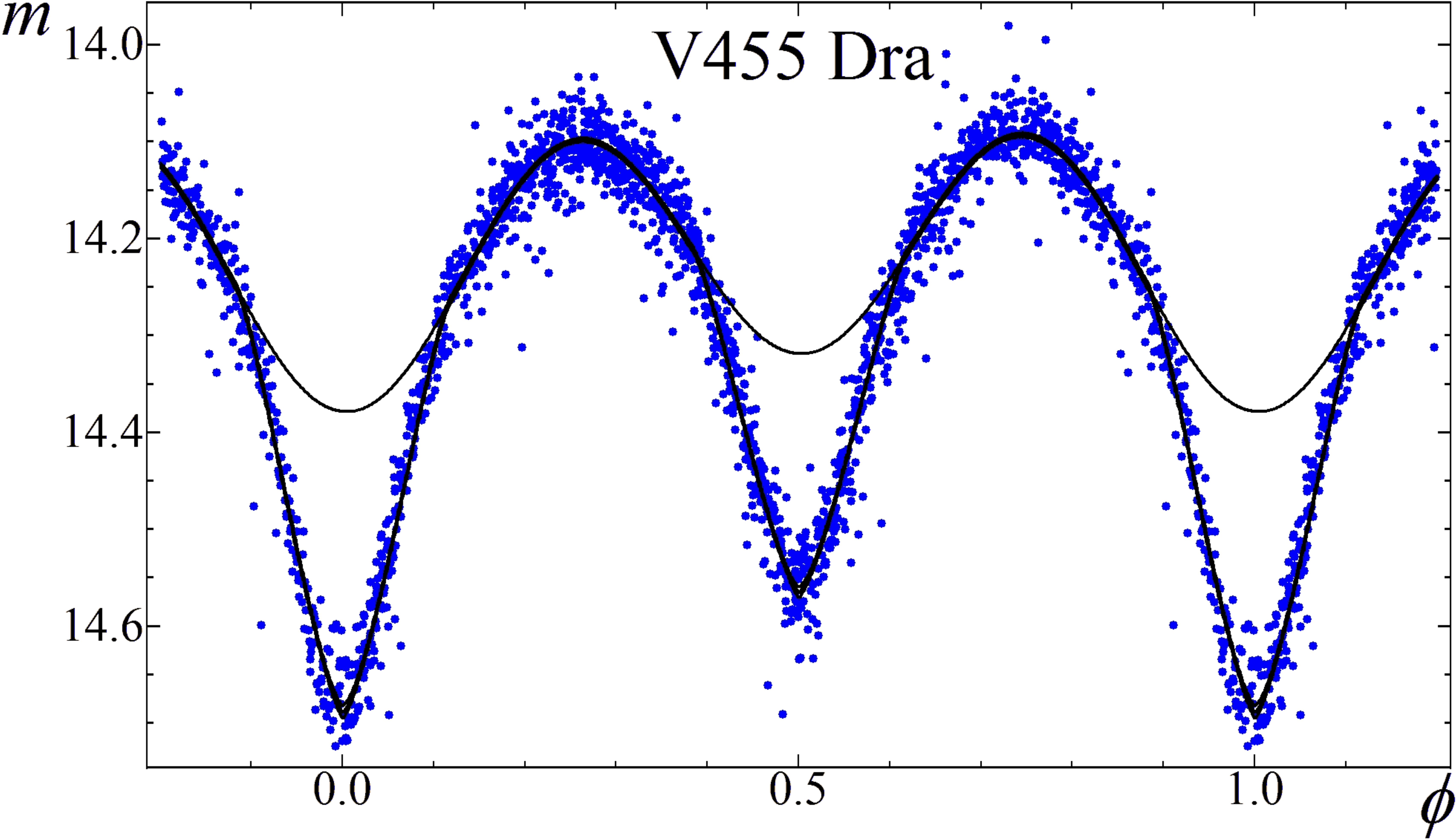}
\caption{The phase light curve of V0455 Dra and it's approximation using the NAV algorithm. A smooth line above the eclipse corresponds to the ''out-of-eclipse'' continuum apptoximated by a trigonometrical polynomial of order 2.\label{fig:jkasfig2}}
\end{figure}

\begin{figure}
\centering
\includegraphics[width=80mm]{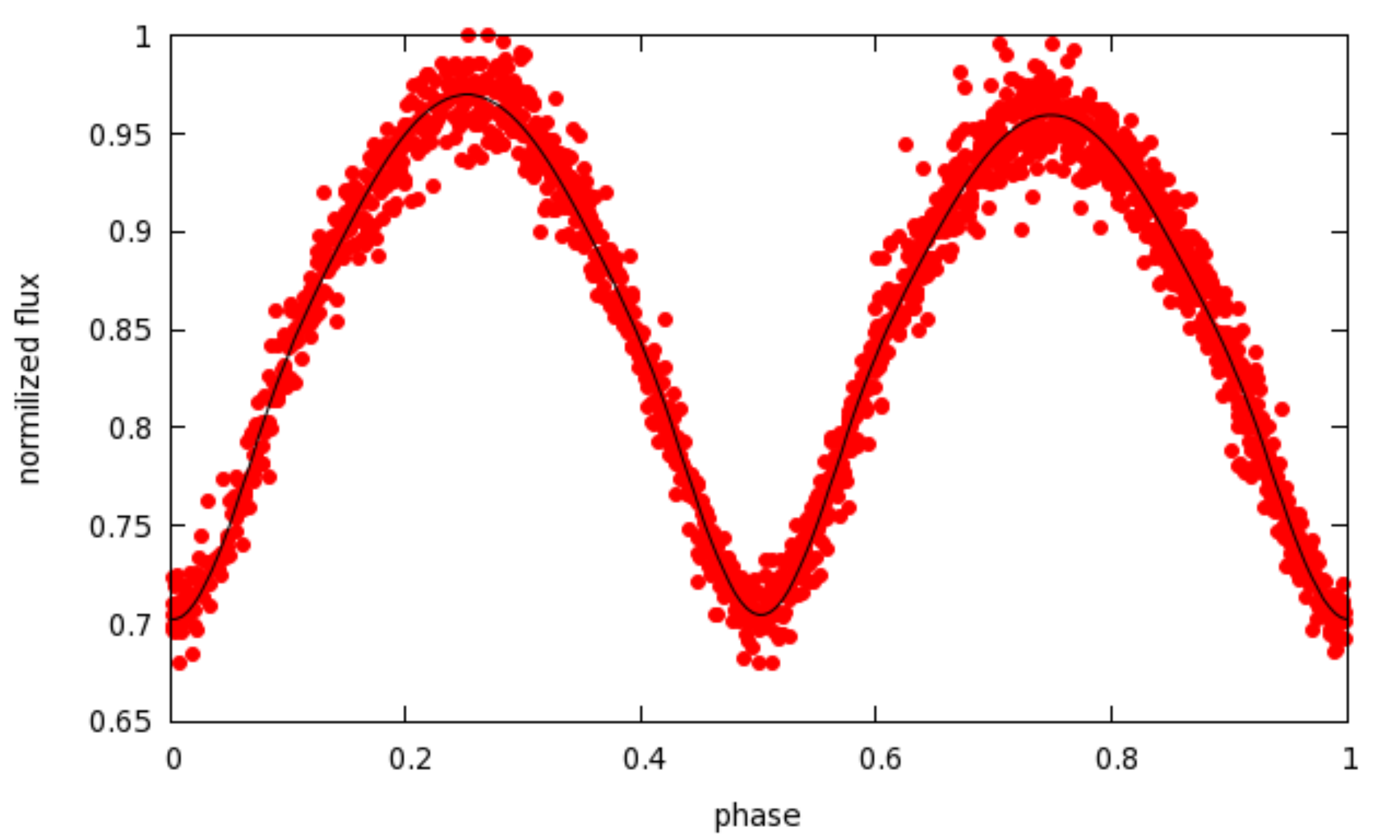}
\caption{The phase light curve of V0454 Dra and it's approximation using Wilson-Devinney algorithm.\label{fig:jkasfig3}}
\end{figure}
\begin{figure}
\centering
\includegraphics[width=80mm]{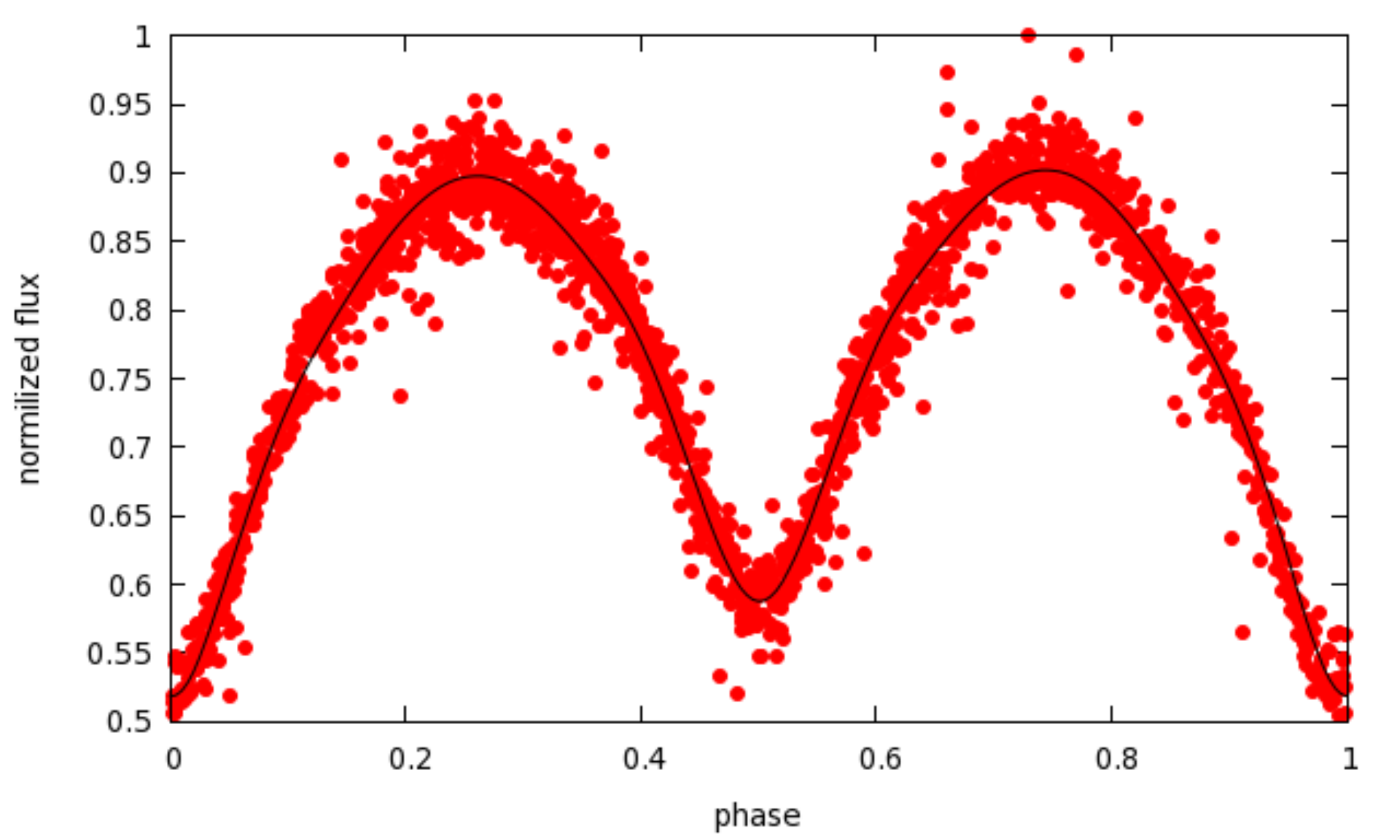}
\caption{The phase light curve of V0455 Dra and it's approximation using Wilson-Devinney algorithm.\label{fig:jkasfig4}}
\end{figure}

\begin{figure}
\centering
\includegraphics[width=80mm]{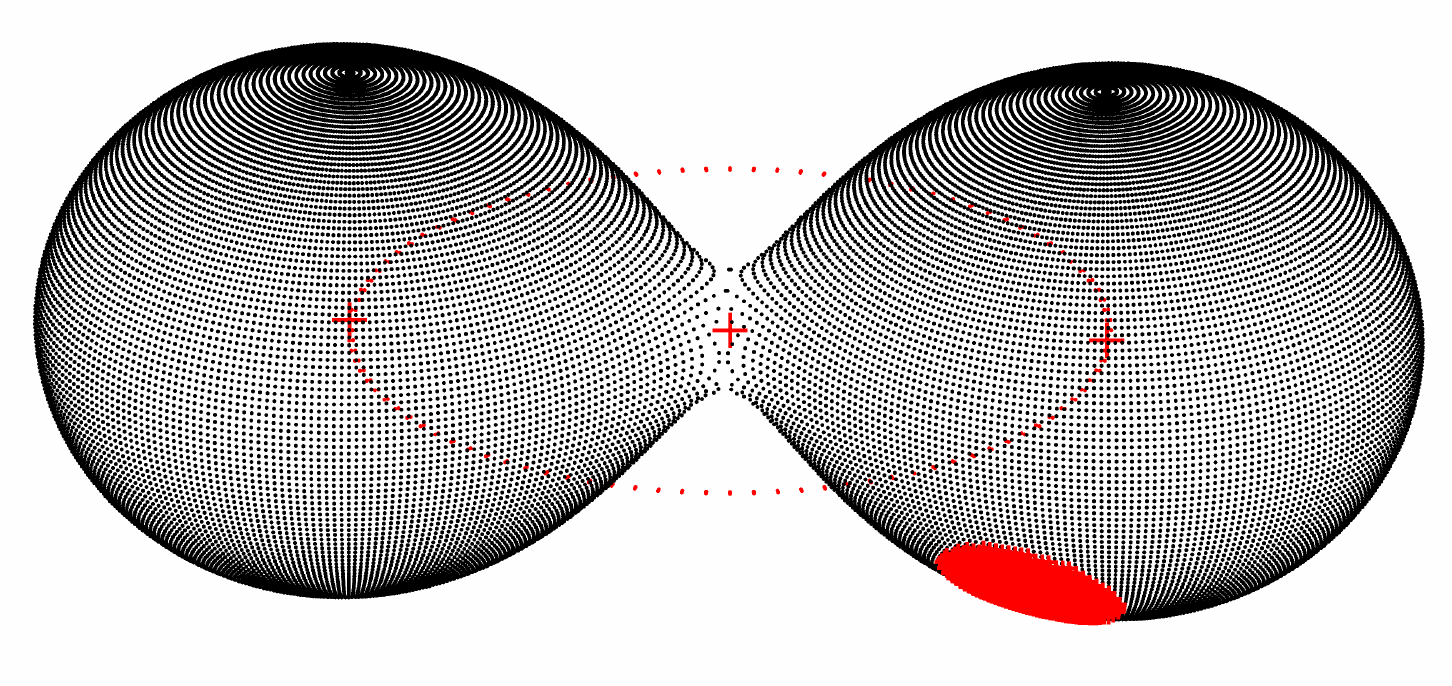}
\caption{The model of V0454 Dra and it's cold spot. The adopted phase for the best viewing is $\phi=0.74.$\label{fig:jkasfig5}}
\end{figure}

\begin{figure}
\centering
\includegraphics[width=80mm]{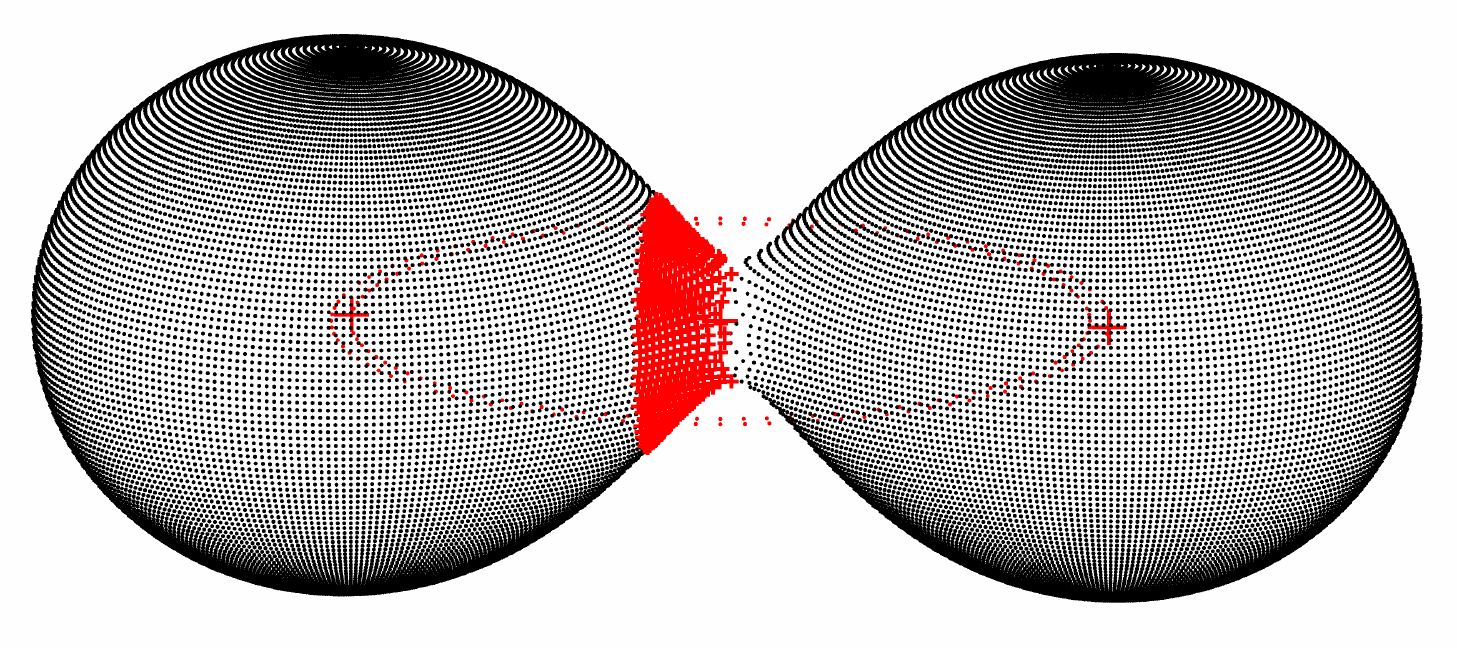}
\caption{The model of V0455 Dra and it's cold spot. The adopted phase for the best viewing is $\phi=0.24.$\label{fig:jkasfig6}}
\end{figure}



\begin{thebibliography}{}

\bibitem[Andronov (1994)]{a94}
Andronov, I.L. (1994)
(Multi-) Frequency Variations of Stars. Some Methods and Results.
Odessa Astronomical Publications, 7, 49

\bibitem[Andronov (2012)]{a12}
Andronov, I. L. (2012)
Phenomenological modeling of the light curves of Algol-type eclipsing binary stars.
Astrophysics, 55, 536

\bibitem[Andronov and Marsakova (2006)]{am06}
Andronov, I.L., Marsakova, V.I. (2006)
Variability of long-period pulsating stars. I. Methods for analyzing observations.
Astrophysics, 49, 370

\bibitem[Andronov et al.(2003)]{a03}
Andronov, I.L., Antoniuk, K.A., Augusto, P., et al.
(2003)
Inter-longitude astronomy project: some results and perspectives.
Astronomical and Astrophysical Transactions, 22, 793

\bibitem[Andronov et al.(2008)]{a08}
Andronov, I.L., Chinarova, L.L., Han, W., et al. (2008)
Multiple timescales in cataclysmic binaries. The low-field magnetic dwarf nova DO Draconis.
Astronomy and Astrophysics, 486, 855


\bibitem[Andronov et al.(2015)]{a15}
Andronov, I.L.; Kim, Yonggi, Kim, Young-Hee, et al. (2015)
Phenomenological Modeling of Newly Discovered Eclipsing Binary 2MASS J18024395 + 4003309 = VSX J180243.9+400331.
Journal of Astronomy and Space Science, 32, 127

\bibitem[Andronov et al. (2017)a]{a17a}
Andronov, I.L., Andrych K.D., Antoniuk, K.A., et al. (2017a)
Instabilities in Interacting Binary Stars.
 ASP Conference Series, 511, 43

\bibitem[Andronov et al.(2017)b]{a17b}
Andronov, I. L., Tkachenko, M.G., Chinarova, L.L. (2017b)
Comparative Analysis of Phenomenological Approximations for the Light Curves of Eclipsing Binary Stars with Additional Parameters. Astrophysics, 60, 57

\bibitem[Andrych et al.(2015)]{aac15} 
Andrych, K.D., Andronov, I.L., Chinarova L.L., Marsakova V.I. (2015) ''Asymptotic Parabola'' Fits for Smoothing Generally Asymmetric Light Curves, Odessa Astronomical Publications,  28, 158

\bibitem[Andrych et al.(2020)]{aac20}
Andrych, K.D., Andronov, I.L., Chinarova L.L. (2020) MAVKA: Program Of Statistically Optimal Determination Of Phenomenological Parameters Of Extrema. Parabolic Spline Algorithm and Analysis of Variability of the Semi-Regular Star Z UMa, Journal of Physical Studies, 24, 1902 

\bibitem[Andrych and Andronov (2019)]{aa19} 
Andrych, K.D., Andronov, I.L. (2019)
MAVKA: Software for statistically optimal determination of extrema
Open European Journal on Variable Stars, 197, 65
2019OEJV..197...65A

\bibitem[Bradstreet (2005)]{bra05}Bradstreet, D.H. (2005) Fundamentals of  Solving Eclipsing Binary Light Curves Using  Binary Maker  3.,  SASS, 24, 23

\bibitem[Gaia DR2 (2018)a]{gaia18a}Gaia Collaboration (2018) VizieR Online Data Catalog: Gaia DR2, 2018yCat.1345....0G

\bibitem[Gaia DR2 (2018)b]{gaia18b}Gaia Collaboration (2018) Gaia Data Release 2. Summary of the contents and survey properties, Astronomy \& Astrophysics, 616, id.A1, 22 pp.

\bibitem[Han et al. (2017)]{k17}
Han, Kiyoung, Kim, Yonggi, Andronov, Ivan L., et al. (2017)
Quasi-Periodic Oscillation of a Magnetic Cataclysmic Variable, DO Draconis.
Journal of Astronomy and Space Sciences, 34, 37

\bibitem[Henden (2007)]{h07}
Henden A., 2007, ftp://ftp.aavso.org/public/calib/dodra.dat

\bibitem[Kazarovets et al.(2015)]{k15}Kazarovets, E.V., Samus, N.N., Durlevich, O.V., Kireeva,  N.N., Pastukhova, E.N. 2015, The 81st Name-List of Variable Stars. Part I - RA 00h to 17h30, IBVS, 6151, 1

\bibitem[Kallrath and Milone (2009)]{km09}
Kallrath, J.,  Milone, E.F. (2009) Eclipsing Binary Stars: Modeling and Analysis.  Springer-Verlag New York, 444p.

\bibitem[Kreiner et al. (2003)]{k01}
Kreiner, J.M., Kim, Chun-Hwey, Nha, Il-Seong (2001)
An Atlas of O-C Diagrams of Eclipsing Binary Stars,  Cracow, Poland: Wydawnictwo Naukowe Akademii Pedagogicznej (2001aocd.book.....K)

\bibitem[Kreiner et al. (2003)]{k03}
Kreiner J.M., Rucinski S., Zola S.  et al. (2003) Physical parameters of components in close binary systems. I, Astron. Astrophys., 412, 465

\bibitem[Marsakova and Andronov (1996)]{ma96}
Marsakova, V.I., Andronov, I.L. (1996)
Local Fits of Signals with Asymptotic Branches.
Odessa Astronomical Publications, 9, 127

\bibitem[Mikul\'a\v{s}ek (2015)]{m15}
Mikul\'a\v{s}ek, Z. (2015)
Phenomenological modelling of eclipsing system light curves
Astronomy \& Astrophysics, 584, id.A8 

\bibitem[Samus et al. (2017)]{s17}Samus, N.N., Kazarovets, E.V., Durlevich, O.V., et al.
2017, General Catalogue of Variable Stars: Version GCVS 5.1, Astronomy Reports, 61, 80

\bibitem[Tkachenko et al. (2016)]{t16}
Tkachenko, M.G., Andronov, I.L., Chinarova, L.L. (2016)
Phenomenological Parameters of the Prototype Eclipsing Binaries Algol, $\beta$ Lyrae and W UMa.
Journal of Physical Studies, 20, 4902
 
\bibitem[Vavilova et al. (2017)]{v17}
 Vavilova, I.B., Yatskiv, Ya. S., Pakuliak, L. K., et al. (2017) UkrVO Astroinformatics Software and Web-services. Proceedings of the International Astronomical Union, 12, 361

\bibitem[Virnina (2010)]{v10}Virnina, N.A. 2010, New Binary Systems With Asymmetric Light Curves,
Odessa Astronomical Publications, 23, 143


\bibitem[Virnina (2011)]{v11}Virnina, N.A. 2011, “Tsessevich” Project: an Attempt to Find the System YY Dra. I, Open European Journal on Variable stars, 133, 1 

\bibitem[Wilson (1979)]{w79}
Wilson R.E. (1979) Eccentric Orbit Generalization and Simultaneous Solution of Binary Star Light and Velocity Curves, ApJ, 234, 1054

\bibitem[Wilson (1994)]{w94}
Wilson R.E. Binary-star light Curve Models (1994) PASP, 106, 921

\bibitem[Wilson and  Devinney (1971)]{wd71}Wilson R.E.,  Devinney E.J. (1971) Realization of Accurate Close-Binary Light Curves: Application to MR Cygni, Astrophys.J., 166, 605 

\bibitem[Zo{\l}a et al. (1997)]{z97}Zo{\l}a, S., Kolonko, M., Szczech, M. (1997) Analysis of a Photoelectric Light Curve of the W UMa-Type Binary ST Ind, A\&A, 324, 1010

\bibitem[Zo{\l}a et al. (2010)]{z10}Zo{\l}a, S., Gazeas, K., Kreiner, J.M., et al. (2010) 
Physical Parameters of Components in Close Binary Systems – VII, MNRAS, 408. –  464




\end{thebibliography}
\end{document}